\documentclass[12pt]{article}
\usepackage{amsmath,amssymb,epsfig}
\usepackage{color}
\input{colordvi.tex}
\def\unit{{\relax{\rm 1\kern-.26em I}}}

\makeatletter

\renewcommand\section{\@startsection {section}{1}{\z@}%
                                   {-3.5ex \@plus -1ex \@minus -.2ex}%
                                   {2.3ex \@plus.2ex}%
                                   {\normalfont\large\bfseries}}

\renewcommand\subsection{\@startsection{subsection}{2}{\z@}%
                                     {-3.25ex\@plus -1ex \@minus -.2ex}%
                                     {1.5ex \@plus .2ex}%
                                     {\normalfont\normalsize\bfseries}}

\newcount\hour \newcount\minute
\hour=\time \divide \hour by 60
\minute=\time
\def\now{%
\ifnum \hour<13
  \ifnum \hour=0 \advance \hour by 12 \number\hour:\else \number\hour:\fi%
     \ifnum \minute<10 0\fi%
     \number\minute%
\ A.M.%
\else \advance \hour by -12 \number\hour:%
  \ifnum \minute<10 0\fi%
  \number\minute%
  \ P.M.%
\fi%
}

\makeatother

\begin{document}

% format
\baselineskip=18pt  % a la harvmac
\numberwithin{equation}{section}  % make eq labels (sec.num)
\allowdisplaybreaks  % allow page breaks in displayed eqs

% print date, time and filename
%\pagestyle{myheadings}
%\markright{{\tt \jobname.tex} -- \today{} \now}

%%%%%%%%%%%%%%%%%%%%%%%%%%%%%%%%%%%%%%%%%%%
%%%        TITLE BEGINS HERE
%%%%%%%%%%%%%%%%%%%%%%%%%%%%%%%%%%%%%%%%%%%

%% ========== title (note version) begins here ==========
%
%\vspace*{-1cm}
%\begin{center}
% {\Large\bf Title of the Document}
%\end{center}
%\vspace*{-.5cm}
%
%% ========== title (note version) ends here ==========

%% ========== title (paper version, a la harvmac) begins here ==========

\thispagestyle{empty}

% Report number
\vspace*{-2cm}
\begin{flushright}
%{\tt arXiv:yymm.nnnn}\\
\end{flushright}

\begin{flushright}
KYUSHU-HET-137
\end{flushright}

\begin{center}

\vspace{1.5cm}

\vspace{1.0cm}
{\bf \Large Discrete Gauge Symmetry and   }
\vspace{0.3cm}

{\bf \Large Aharonov-Bohm Radiation in String Theory}
\vspace*{0.2cm}

\vspace{1.3cm}

{\bf
Yutaka Ookouchi}
\vspace*{0.5cm}

{\it Faculty of Arts and Science, Kyushu University, Fukuoka 819-0395, Japan  }\\

\vspace*{0.5cm}

\end{center}

\vspace{1cm} \centerline{\bf Abstract} \vspace*{0.5cm}

We investigate cosmological constraints on phenomenological models with discrete gauge symmetries by discussing the radiation of standard model particles from Aharonov-Bohm strings. Using intersecting D-brane models in Type IIA string theory, we demonstrate that Aharonov-Bohm radiation, when combined with cosmological observations, imposes constraints on the compactification scales.

\newpage
\setcounter{page}{1} % don't number title page

%% ========== title (paper version, a la harvmac) ends here ==========

%%%%%%%%%%%%%%%%%%%%%%%%%%%%%%%%%%%%%%%%%%%
%%%           TITLE ENDS HERE
%%%%%%%%%%%%%%%%%%%%%%%%%%%%%%%%%%%%%%%%%%%

%%%%%%%%%%%%%%%%
\section{Introduction}

A discrete symmetry is one of the key ideas in constructing a phenomenologically viable model of particle physics. Various applications have been discussed (see \cite{Yanagida,examples,Ishimori} and references therein) in the contexts of stabilizing baryons, suppressing flavor changing neutral currents and fixing the quark and lepton mixing angles. Recent arguments by Banks and Seiberg shed new light on this subject from a different viewpoint \cite{BS}. It has been known that no continuous global symmetry is allowed in a consistent quantum theory of gravity such as superstring theories \cite{OldNoGlobal,NewNoGlobal}. Banks and Seiberg extended this ``no global symmetries'' theorem to also include discrete symmetries. Then, their arguments imply that, when one engineers phenomenological models in string theories, such discrete symmetries have to be gauged (if it is not broken explicitly). One of the typical origins of discrete gauge symmetries is an extra $U(1)$ symmetry which is ubiquitous in various standard model-like theories engineered in string theories. Such a gauge symmetry is broken due to BF coupling (or St\"uckelberg coupling), but there often exists a remnant discrete gauge symmetry. In this sense, the appearance of discrete gauge symmetries is generic in string phenomenology.

The universal effective Lagrangian description of $ \mathbb{Z}_k$ discrete gauge theory was initially studied in \cite{OldAB} and recently revisited in \cite{BS} with a streamlined argument in light of BF coupling. In the effective theory, there are several observables characterizing the discrete gauge theory. One of them is an Aharonov-Bohm (AB) string \cite{OldAB}. A characteristic feature of the AB string is the ``confined magnetic flux''. In $ \mathbb{Z}_k$ theory, it has $1/k$ unit of fundamental magnetic charge. Because of it, the Aharonov-Bohm string can nontrivially interact with an Aharonov-Bohm particle which is another observable characterizing the theory and carries an electric charge smaller than $k$ in the fundamental unit. By circling around the string, the AB particle picks up a non-trivial holonomy. This is similar to the well-known Aharonov-Bohm effect of a solenoid \cite{Original}. This effect can be viewed as an interaction of the gauge potential $A_{\mu}$ created by the string with the current ${\cal J}^{\mu}$ constructed out of the particle, ${\cal L}\sim A_{\mu}{\cal J}^{\mu}$.

In string theories, discrete gauge symmetries have been widely studied from various points of view \cite{DSmany,DGS1,DGS2,DGS3,DGS4,DGS5}. For example, in \cite{DGS4}, BF coupling is derived by dimensionally reducing the Chern-Simons coupling in a flux compactification, and, in \cite{DGS1}, by dimensional reduction of the kinetic term of an RR-field in a compactification with discrete torsions. It is fascinating that seemingly different such studies can be understood in a uniform manner by BF couplings. An interesting aspect of discrete gauge symmetries realized in string theories is that AB strings and AB particles can be dynamical objects even if they are probes in the field theory limit. In this paper, we discuss such dynamical AB strings/particles in string theory and make a connection to cosmological observations through the Aharonov-Bohm interaction.  

To achieve this goal, the key ingredient is to figure out the radiation of AB particles from AB strings, first pointed out in \cite{AW}. Recently, the radiation of bosonic and fermionic particles from moving solenoids has been explicitly calculated \cite{AB1,AB2}. Based on this interesting progress, we would like to go a step further, especially in the direction of understanding AB strings realized in string theories. It is worthy emphasizing that AB particles realized in string theory setup can be electrons or quarks, so AB radiation in this case is direct emission of particles in the standard model. Also in string theories, the existence of the extra dimensions, which makes a reconnection probability of strings small, increases the production rate of the AB particles.

The outline of this paper is as follows. In section 2, we first show one of the examples of discrete gauge symmetries realized in string theories. Then, we investigate the associated AB strings/particles in the context of D-branes and fundamental strings\footnote{See \cite{cosmicsuperstring} for earlier works on cosmic superstrings and \cite{reviewsuperstring} for reviews.}. In section 3, we review the explicit calculations of the power of AB radiation along the lines of \cite{AB1,AB2}, and discuss the formulae that can be applicable to our models in a wide range of the parameter space. Then, we compute the total power of radiation from various sizes of loops. In section 4, we impose cosmological constraints arising from the Big Bang Nucleosynthesis (BBN) and the diffuse $\gamma$-ray background. In appendix A, we exhibit an example of standard model-like theories in Type IIA theory by means of intersecting D6-branes \cite{DGS2}.

%%%%%%%%%%%%%%%%%%%%%%%%%%%%%%%%
\section{ $ \mathbb{Z}_k$ discrete gauge symmetry in Type IIA theory}

In this section, we introduce a simple example of  D-brane models with $\mathbb{Z}_k$ discrete gauge symmetry, and study the energy scales of AB strings and AB particles. Since constructing a realistic standard model-like theory is technically involved and the details of the construction are not relevant in the present context, here we study a simplified version of the models and extract the essential phenomenon. An illustrating example of intersecting D6-brane models is shown in appendix A. Although we focus on a D-brane model for concreteness, which is one of recent progresses on discrete gauge symmetries in string theories \cite{DGS1,DGS2,DGS3,DGS4}, the essential arguments can be applied to various related models. 

Consider Type IIA string theory compactified on a Calabi-Yau manifold with a pair of three-cycles intersecting once with each other $(\alpha \cdot \beta)=1$. Suppose a D6-brane is wrapping on the cycle $\beta$. In this case, the reduction of the Chern-Simons couplings in the D6-brane action leads to the following BF coupling \cite{DGS2}
\begin{equation}
k \int_{4D } \widetilde{C}_2\wedge dA  ,\qquad {\rm where  } \qquad \widetilde{C}_2=\int_{\beta}C_5, 
\end{equation}
where $k$ is the wrapping number and $C_5$ is the Ramond-Ramond five-form field. This BF-coupling indicates the existence of $ \mathbb{Z}_k$ gauge symmetry in the four-dimensional effective theory. At first sight, the action seems to preserve continuous $U(1)$ symmetry. However, from the gauge transformations of the dual fields\footnote{For example, the dual scalar is defined by $d\widetilde{C}_2=*_{4}d\phi$ with $\phi=\int_{\alpha}C_3$.  } $\phi$ and $V_{\mu}$, one can observe non-linear transformations which imply the breaking of $U(1)$ symmetry,
\begin{eqnarray}
&A \to A +d \lambda,\quad & \phi \to \phi +k \lambda , \nonumber \\
&\widetilde{C}_2\to \widetilde{C}_2 +d \Lambda ,\quad  & V \to V+k\Lambda . \nonumber
\end{eqnarray}
Following \cite{BS}, let us consider Aharonov-Bohm strings/particles in this theory. AB particle is an object which electrically couples to the gauge field on the D6-brane. It is represented as a line operator by using a closed world-line (or infinite length of world-line) $\Sigma_1$,
\begin{equation}
{\cal O}_{\rm AB\ particle}\sim {\rm exp}\left( {i\int_{\Sigma_1}A} \right). 
\end{equation}
Hereafter, we assume that AB particles carry the minimum charge of the discrete symmetry.
From a ten-dimensional point of view, such an object is identified with the fundamental string ending on the D6-brane. $k$ periodicity can be seen as a fact that $k$ AB particles are annihilated with an instanton. With the dual field $\phi$, one can consider an instanton operator $e^{-i\phi}$. However, this operator is not gauge invariant. To compensate the gauge non-invariance, we add line operators, $ {\rm exp}[ {-i \phi +ik \int_{L}A} ]$. $L$ is the world-line ending at the point $P$ where the instanton operator is inserted. Boundary contribution from the added line operator makes the total operator gauge invariant. The $k$ world-lines disappear at the point $P$, and this phenomenon can be understood that the instanton annihilates $k$ AB particles. This instanton operator electrically couples to $\phi$, thus magnetically couples to $\widetilde{C}_2$. In ten-dimension, it is interpreted as a D2-brane wrapping on the cycle $\alpha$.

An AB string is a string-like object electrically couples to $\widetilde{C}_2$. The operator corresponding to the AB string is written as follows:
\begin{equation}
{\cal O}_{\rm AB \ string}\sim {\rm exp} \left( i \int_{\Sigma_2} \widetilde{C}_2   \right),
\end{equation}
where $\Sigma_2$ is a closed surface (or infinitely large world-sheet). In string theory, such an object couples electrically to $C_5$, and is interpreted as a D4-brane wrapping on the cycle $\alpha$. $k$ AB strings annihilate with the monopole operator $e^{-i \int_{L}V}$ (or junction operator in the terminology used in \cite{DGS4}). Consider a world-sheet $C$ with boundary $L$. Following the same argument above, it is easy to check that the operator, $  {\rm exp} [ {-i \int_{L}V +ik \int_{C}\widetilde{C}_2} ]$, is gauge invariant. Since the monopole is electrically coupled to $V_{\mu}$, or equivalently, magnetically coupled to $A_{\mu}$, we can identify it with a D4-brane wrapping on the $\beta$ cycle and ending on the D6-brane.

The energy scales of the AB string and the AB particle highly depend on geometrical aspects of the D-branes and compactification manifolds. For example, as mentioned above, the AB particle is the fundamental string ending on the D6-brane. If there is no another D6-brane, then the other end of the string has to go to the spatial infinity, which implies that the AB particle has infinite mass and should be interpreted as a probe particle in the field theory. However, in the realistic model shown in appendix A, there is another intersecting D6-brane. In this case, the other end of the fundamental string can be placed on the other brane. Hence, the length of the string can be zero when it is placed at an intersection point of the D6-branes, thus the AB particle is massless\footnote{Eventually they acquire masses via Higgs mechanism and Yukawa interactions.}. Interestingly, in the realistic model, the fundamental strings stretching between two intersecting D6-branes correspond to the particles in the standard model such as leptons and quarks, so we conclude that AB particles are the standard model particles. 

As for the AB string, the tension is governed by the volume $V_3$ of the cycle $\alpha$
\begin{equation}
\mu \equiv T_{D4}V_3= {V_3 \over (2\pi)^4g_s l_s^5}.
\end{equation}
where $T_{D4}$ is the tension of the D4-brane and $g_s$, $l_s$ are the string coupling and length, respectively. An interesting feature of the intersecting D-brane models is the hierarchy between the energy scale of string tension and the mass scale of the AB particle. Because of this fact, Aharonov-Bohm radiation, which we will show below, occurs at a drastic rate, and a large number of the standard model particles are emitted from the AB string. Also, the negligibly small mass of the AB particle makes calculation of AB radiation simple, and allows us to apply the formulae shown in \cite{AB1,AB2} directly to the present case.

The AB string and the AB particle have nontrivial topological interaction. Suppose the particle with the minimum charge circles around the string. Since we put the string in the space-time, the action should include a source term of $C_5$, 
\begin{equation}
k \int_{4D } \widetilde{C}_2\wedge dA  + \int_{\Sigma_2 } \widetilde{C}_2 ,
\end{equation}
where we assumed the minimum flux of the string. The holonomy picked up by the particle is given by
\begin{equation}
{\rm hol}(\gamma)\equiv \exp \left( \int_{\gamma} A \right)=\exp \left( \int_{\sigma } F\right) =\exp \left( {2\pi i \over k} \right) \equiv \exp( i \Phi),
\end{equation}
where $\partial \sigma =\gamma$ and $\Phi$ is the total magnetic flux in the AB string. In the last equality, we used the equation of motion for $C_5$. This phenomenon is the well-known Aharonov-Bohm effect \cite{Original}. The interaction can be represented as a coupling of the gauge potential around the AB string to the current constructed out of the AB particles
\begin{equation}
\int_{4D} A \wedge *_4  {\cal J} \label{ABparticle}.
\end{equation}
In the next section, we will calculate the emission rate of AB particles with this interaction, basically following the paper \cite{AB1,AB2}.

%%%%%%%%%%%%%%%%%
\section{Aharonov-Bohm radiation in string theory}

In this section, we calculate the total power of radiated particles from Aharonov-Bohm strings. The authors of \cite{AB1,AB2} studied radiation of bosons and fermions\footnote{
Calculation of fermionic radiation is slightly involved due to the spin orientations. } from an oscillating string and loops with cusps or kinks within the validity of the wire approximation. Since we estimate rough order of cosmological constraints for such radiation, we will not carefully treat the order one coefficients of the results. In this case, there is no crucial difference between bosonic and fermionic radiation. In the discrete gauge theory, the gauge field around the string is massive, so the situation is slightly different from the solenoid in electromagnetism. Nevertheless, as we will show below, we can simply apply the analysis of \cite{AB1,AB2} to the present case in a wide range of the parameter space.

The first step for the calculation of the radiation power is to determine the gauge potential around the AB string. From the equation of motion for $\widetilde{C}_2$, we obtain 
\begin{equation}
k\partial_{\mu }A_{\nu} = \tilde{J}_{\mu \nu},
\end{equation}
where we ignored the term coming from the kinetic term since it does not play any role below. $\tilde{J}_{\mu \nu}$ is the dual of the string current which can be represented in the wire approximation as follows:
\begin{equation}
\tilde{J}_{\mu \nu}= \epsilon_{\mu \nu \alpha \beta} J^{\alpha \beta}, \nonumber
\end{equation}
\begin{equation}
J^{\alpha \beta}= \int d \tau d \sigma (\dot{X}^{\alpha}{X^{\prime}}^{\beta}-\dot{X}^{\beta} {X^{\prime}}^{\alpha} ) \delta^{(4)} (x-X(\sigma , \tau)), \nonumber 
\end{equation}
where $\sigma, \tau$ are the world-sheet coordinates of the AB string. In the momentum space, the solution of the equation is given by
\begin{equation}
A_{\nu}={1\over k} \epsilon_{\mu \nu \alpha \beta}  {p^{\mu} \over (p_{\lambda}p^{\lambda} ) } J^{\alpha \beta}.
\end{equation}
This is exactly the same as the gauge potential around the solenoid shown in \cite{AW, AB1,AB2}. With this gauge potential and the interaction \eqref{ABparticle}, one can calculate the power of particle emission, following the strategy used in \cite{AB1,AB2}. In the example shown in the appendix, AB particles are leptons and quarks, so the currents should be represented as
\begin{equation}
 {\cal J}_{\mu}=\sum_a \bar{\psi}_a \gamma_{\mu} \psi_a  .   
\end{equation}
The relevant quantity to calculate in our purpose is radiation power from AB string loops with cusps or kinks. For a loop with the size $L$, we naively expect that the typical energy scale is ${\cal O}(1/L)$. However, near the cusps or kinks, high frequency modes should be included. The maximum mode can be understood from the breakdown scale of the wire approximation, namely $N_{\rm max}\sim \sqrt{\mu}L$ where $\mu$ is the tension of the string. According to the concrete calculations in \cite{AB1,AB2}, the total power of massless particle emissions is proportional to $N_{\rm max}$,
\begin{equation}
P_{\rm AB}\simeq \Gamma_{\rm AB} { \Phi^2 \over  L^2} N_{\rm max}=\Gamma_{\rm AB} { \Phi^2 \sqrt{\mu}  \over L}. \label{powerAB}
\end{equation}
$\Gamma_{\rm AB}$ depends on dynamics of the strings and includes the average number of cusps (kinks) per oscillation. Although this formula is reliable for massless radiation, it can be applied to our model in a wide range of the parameter space: The dominant contribution of radiation comes from the high frequency modes $N_{\rm max}\sim \sqrt{\mu} L$, which can be large enough when the loops size is  much larger than the width of the string. From \eqref{powerAB}, we naively expect that the dominant radiated energy is ${\cal O}(\sqrt{N_{\rm max}}/L)$. If this energy scale is much larger than the mass scales of AB particles, then we can treat them as massless particles. By assuming that the string tension is in intermediate scale, this condition is easily satisfied. Also, in this case, it is reasonable to  assume that radiated particles are relativistic.

The next ingredient we should understand for a cosmological consequence of AB radiation is the number density of loops in the universe. In addition to the AB radiation, string loops emit gravitational waves through their oscillations. Its power is well known \cite{VS}, 
\begin{equation}
P_{\rm GW} = \Gamma_{\rm GW} \,G \mu^2 \label{powerGW}.
\end{equation}
$G$ is the Newton constant and $\Gamma_{\rm GW} = {\cal O}(10)$ depends on dynamics of strings (for example see \cite{VS}). Comparing \eqref{powerAB} with \eqref{powerGW}, we see that Aharonov-Bohm radiation is dominant for loops with small size. The critical length is 
\begin{equation}
L_{\rm crit}={\Gamma_{\rm AB} \Phi^2 \over \Gamma_{\rm GW}} \left({1\over G\mu} \right)^{3/2} \sqrt{G} .
\end{equation}

As in the standard cosmic strings \cite{VS}, AB strings are also assumed to reach the scaling regime by constantly loosing the energy by the gravitational wave and the AB radiation. In this assumption, the number density of the loop size between $L $ and $L+d L$ is given by 
\begin{equation}
n_L dL ={\xi \sqrt{\alpha}   \over p^{b} L^{5/2} t^{3/2} } \label{number density},
\end{equation}
in the radiation-dominated era \cite{VS,Vach,Dufaux,dilaton}. Here, we added the effect of the reconnection probability. In string theories, due to the extra dimensions, the reconnection probability is smaller than unity \cite{JonsPol}. However, the exponent $b>0$ is ambiguous, which is usually expected to be one \cite{DV1}, but in some cases, it can be larger or smaller than one \cite{Tye1,AvS}. Also, the loop number density depends crucially on the size of a loop produced from a long string. This is the maximum size of the loops. According to the recent computer simulations of string networks \cite{sim1, sim2}, the typical initial size of the loops is governed by the cosmic time $t$ when they are produced, $L_{\rm max}\simeq \alpha t$ with $\alpha \sim 0.1$.  The parameter $\xi $ is also determined by dynamics of cosmic string network. However, since there is no computer simulation of string network for AB strings, we simply speculate that it is the same order $\xi={\cal O}(10)$ as the one for the standard cosmic strings \cite{VS}.

\begin{figure}[htbp]
\begin{center}
 \includegraphics[width=.4\linewidth]{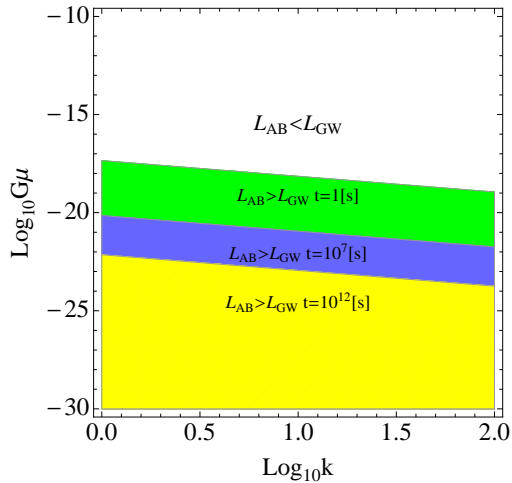} 
\vspace{-.1cm}
\caption{\sl The vertical and horizontal axes are $\log_{10} G\mu$ and $\log_{10} k$ respectively. In the white region, the condition $L_{\rm AB}<L_{\rm GW}$ is satisfied for $\Gamma_{AB}=\Gamma_{\rm GW}=50$. The green, blue and yellow regions satisfy $L_{\rm AB}> L_{\rm GW}$ at $t=1, 10^7 , 10^{12}${\rm [s]} respectively.  }
\label{shortest}
\end{center}
\end{figure}

Now we are ready to calculate the total power of the Aharonov-Bohm radiation. The power radiated by all loops is 
\begin{equation}
\dot{\rho}=\int_{L_{\rm min}}^{L_{\rm max}} dL \, n_L P_{\rm AB}.
\end{equation}
The shortest size of the loops, $L_{\rm min}$, is determined by competition of the two types of radiation: If the life-time of a loop is shorter than one Hubble time, then it cannot be observed in the universe. The power radiated by the gravitational wave is given by \eqref{powerGW}. So the condition for the shorter life-time is given by $P^{-1}_{\rm GW}\mu L < t$, 
\begin{equation}
L< \Gamma_{\rm GW} G\mu t \equiv L_{\rm GW}.
\end{equation}
In the same way, we estimate the contribution from the Aharonov-Bohm radiation, $P^{-1}_{\rm AB} \mu L< t$, 
\begin{equation}
 L^2< t {\Gamma_{\rm AB} \Phi^2 \over \sqrt{\mu}}\equiv L_{\rm AB}^2.
\end{equation}
Hence, the minimum size of the loop is written as 
\begin{equation}
L_{\rm min}\equiv {\rm max}[ L_{\rm GW},\  L_{\rm AB}]. 
\end{equation}
Since the two lengths scale as $L_{\rm GW}\sim t$, $L_{\rm AB}\sim \sqrt{t}$, in the early stage the Aharonov-Bohm radiation is dominant. In the figure \ref{shortest}, we show the regions satisfying $L_{\rm AB}>L_{\rm GW}$ at $t=1, 10^7, 10^{12} {\rm [s]}$.

 %%%%%%%%%%%%
%%%%%%%%%%%%

%%%%%%%%%
\section{Cosmological constraints}

In this section, we discuss cosmological constraints arising from the Big Bang Nucleosynthesis (BBN) and the diffuse $\gamma$-ray background. In addition to these constraints, we have also investigated constraints coming from the Cosmic Microwave Background (CMB), re-ionization and the effective number of neutrinos along the lines of \cite{Raxion1,Nakayama}\footnote{We would like to thank K. Kamada for useful conversations and sharing his unpublished notes. } (see \cite{Kawasaki} for a review). But since these constraints are relatively weak and covered by the BBN and the diffuse $\gamma$-ray background, we do not exhibit them here. 

In a field theory, topological solitons can be usually formed by the Kibble-Zurek mechanism \cite{Kibble} in the expanding universe. However, in the present string model, the phase transition corresponds to compactification of the internal space. To construct the standard model-like theories, we assume the existence of some of D6-branes. At the same time, if there is a few anti D6-branes floating in the spacetime, it is natural to guess that there also exist lower dimensional branes generated by D6-brane/anti D6-brane annihilation \cite{GK}. In any case, here we assume the existence of wrapped D4-branes corresponding to the Aharonov-Bohm strings in the early stage of the universe.

\subsection{Big Bang Nucleosynthesis}

To impose the BBN constraint, it is useful to introduce the energy injected in one Hubble time, $\rho_{\rm AB} \equiv t \dot{\rho}$, and divide it by the entropy density, $s=(g_*/45\pi)^{1/4}(M_{\rm pl}/t)^{3/2}/4$, 
\begin{equation}
{\rho_{AB} \over s}\simeq 1.1\times 10^{-45}     \left({10.75 \over g_*} \right)^{1/ 4}\left(  {t \over 1 {\rm [s]}}  \right)^{5/2}\left(  {\hbar \over 6.5\times 10^{-16} {\rm [eV\cdot s]}}  \right)^{3/ 2} {\rm [GeV]} \left[ {1 {\rm [s]^4}\over M_{\rm pl}} \dot{\rho} \right],
\end{equation}
where $M_{\rm pl}$ is the Plank scale, $M_{\rm pl}=G^{-1/2}\simeq1.2\times 10^{19}$ [GeV]. Using \eqref{powerAB} and \eqref{number density}, we obtain the total power of the Aharonov-Bohm radiation,
\begin{eqnarray}
{1 {\rm [s]}^4 \over M_{\rm pl}}   \dot{\rho}&=&{1 {\rm [s]}^4 \over M_{\rm pl}}  \int^{L_{\rm max}}_{L_{\rm min}} d L\, n_L P_{\rm AB} = {1 {\rm [s]}^4 \over M_{\rm pl}}{2\Phi^2  \xi \sqrt{\alpha} \sqrt{\mu} \over 5p^b t^{3/2}}  \Gamma_{\rm AB} L_{\rm min}^{-5/2}+\cdots \\
&\simeq     & {2\xi \sqrt{\alpha}\over 5 p^b } \times  \left\{
\begin{array}{ll}
{\Phi^2\Gamma_{\rm AB}  \over  \Gamma_{\rm GW}^{5/2}} \left(  {1 {\rm [s]} \over t}  \right)^{4} (G\mu)^{-2} &\quad \text{for}\quad L_{\rm min}=L_{\rm GW} \\ 
1.1\times 10^{54} \ ( \Phi^{ 2}\Gamma_{AB})^{-1/ 4} \left({ 1 \text{[s]}\over t} \right)^{{11/ 4}}(G\mu)^{9/ 8} & \quad \text{for}\quad L_{\rm min}=L_{\rm AB} .
 \end{array}\right. \nonumber
\end{eqnarray}
For convenience below, we also write down $\rho_{\rm AB}/s$, 
\begin{eqnarray}
{\rho_{AB} \over s} \simeq \left\{
\begin{array}{ll}
4.6\times 10^{-46} \left(  { 1 {\rm [s]}\over t}  \right)^{{3/ 2}} \left(  {10.75 \over g_*}  \right)^{1/ 4} { \xi \sqrt{\alpha}   \over p^b } {\Phi^2    \Gamma_{\rm AB} \over  \Gamma_{\rm GW}^{5/2}} (G\mu)^{-2} \ {\rm [GeV]}   &  \ \text{for}\  L_{\rm min}=L_{\rm GW}  \\ 
5.5 \times 10^{8} \left(  {1 {\rm [s]}\over t}  \right)^{{1/ 4}}\left(  {10.75 \over g_*}  \right)^{1/ 4}  { \xi \sqrt{\alpha}   \over p^b }  ( \Phi^{2} \Gamma_{\rm AB})^{-1/4}  (G\mu)^{9/8}\  {\rm [GeV]}   &  \ \text{for}\ L_{\rm min}=L_{\rm AB}. \label{result}
 \end{array}\right. 
\end{eqnarray}
A remarkable feature is the negative power of $G\mu$ for the case $L_{\rm min}=L_{\rm GW}$. As the tension of the string decreases, the total radiated power increases. Hence, naively there is a chance to obtain the lower bound of the string tension, which is complementary to the future gravitational wave experiment. However, the story is not so simple. As shown in figure 1, in the region of small tension, the minimum size of the loop is governed by the AB radiation, $L_{\rm min}=L_{\rm AB}$. In this case, as one can see in the second line of \eqref{result}, $G\mu$ has the positive power, which implies that the total power decreases as the tension becomes small. Owing to this involved fact, the cosmological constraints, which we will show below, depend sensitively on parameter choice.

To begin with, suppose the AB particle is a lepton of the standard model and the AB string associated with the discrete gauge symmetry radiates mainly leptons\footnote{AB strings associated with $ \mathbb{Z}_3^{I}$ symmetry in appendix A are one of examples of this type. $L$, $E_R$ and $N_R$ are emitted from the strings. }. Radiation of leptons after the BBN epoch may cause photo-dissociation process of light elements and changes the light elements abundance. To avoid that, the radiated power should be constrained by \cite{BBN},
\begin{align}
\label{constraint}
\frac{\rho}{s}<\left\{
\begin{array}{ll}
10^{-8}\left({t\over 10^{4} {\rm [s]} } \right)^{-2}\,{\rm [GeV]} & {\rm for} \quad 10^4 {\rm [s]}  \le t < 10^{7} {\rm [s]}, \\
10^{-14}\, {\rm [GeV]} &{\rm for}  \quad 10^7{\rm [s]}  \le t \le 10^{13} {\rm [s]}.\\ 
\end{array}\right.  
\end{align}
%%%%%%%%%%%%
\begin{figure}[htbp]
\begin{center}
 \includegraphics[width=.4\linewidth]{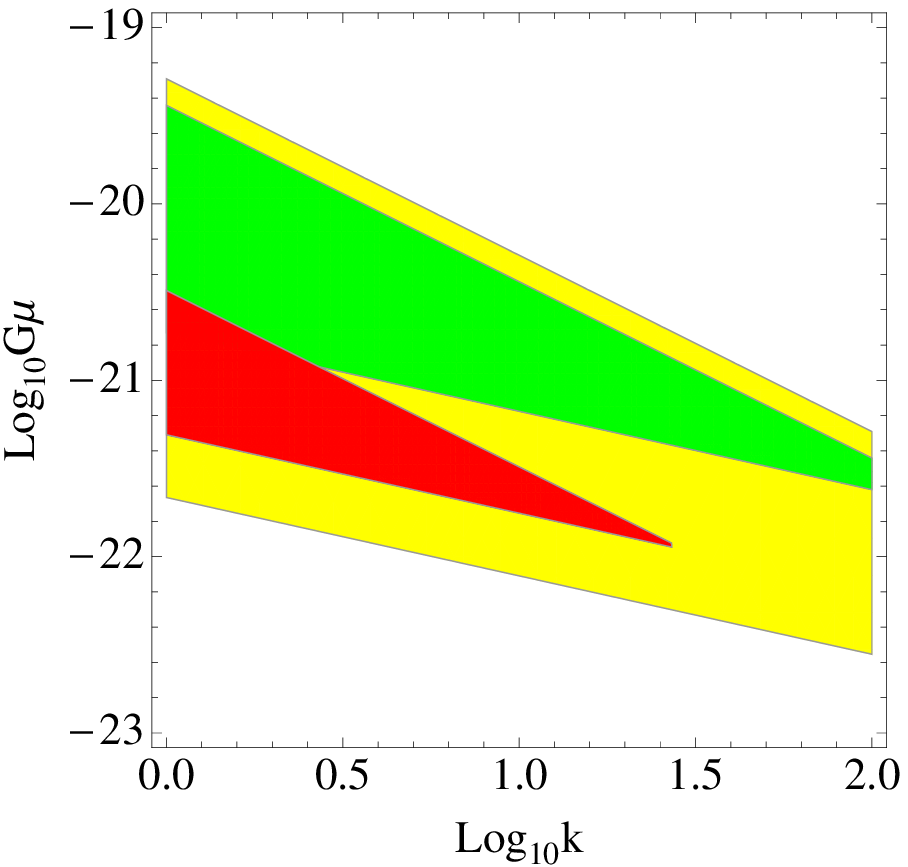} 
 \hspace{3mm}
 \includegraphics[width=.4\linewidth]{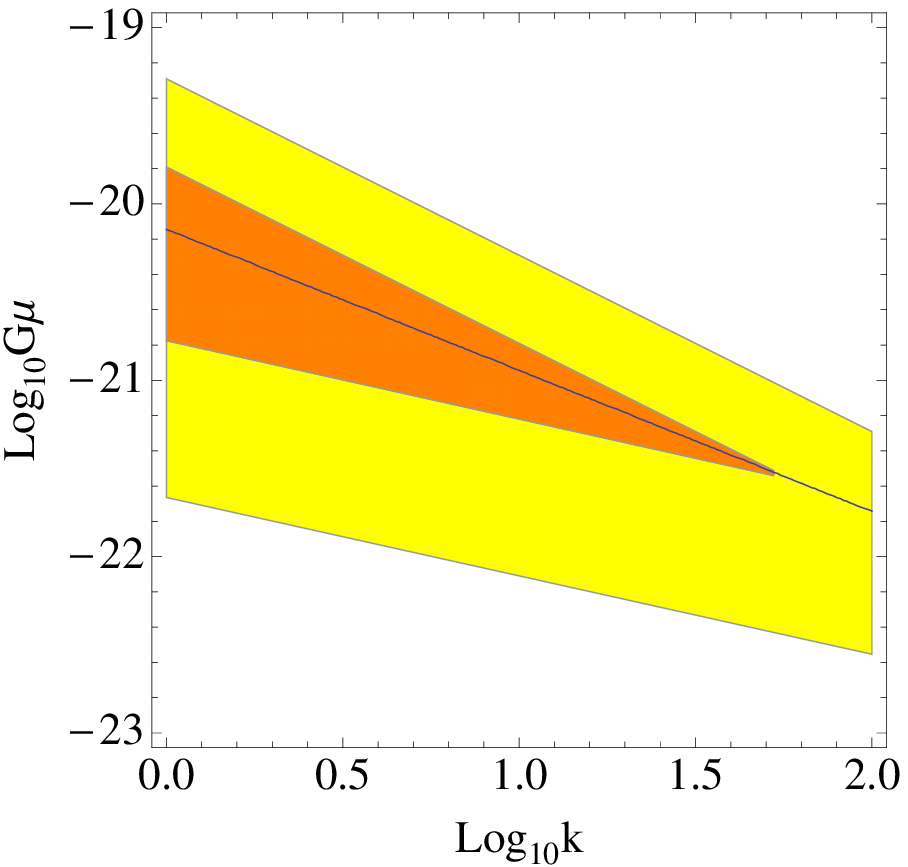} 
\vspace{-.1cm}
\caption{\sl Constraints from leptonic radiation. In the left panel, the green, yellow and red regions are excluded by the conditions \eqref{constraint} for $p^b=10^{-3}$ at $t=10^{6.4}{\rm [s]}$, $10^7{\rm [s]}$ and $10^{8.6}{\rm [s]}$ respectively. The white regions is allowed. We took the parameters, $\alpha=0.1$, $g_*=10.75$, $\Gamma_{\rm AB}=\Gamma_{\rm GW}=\xi=50$. In the right panel, we show two excluded regions at $t=10^7{\rm [s]}$ for $p^b =10^{-2}$ (orange) and $p^b =10^{-3}$ (yellow). The blue line corresponds to $L_{\rm GW}=L_{\rm AB}$ at $t=10^7{\rm [s]}$.}
\label{lepton}
\end{center}
\end{figure}
%%%%%%%%%%%%
Since the condition $L_{\rm min}=\text{max}[L_{\rm GW}, L_{\rm AB}]$ depends on time, constraints become slightly involved. As an illustration, in figure \ref{lepton} we show excluded regions for the parameter choice, $\alpha=0.1$, $g_*=10.75$, $\xi=\Gamma_{\rm GW}=\Gamma_{\rm AB}=50$. From the left panel of the figure \ref{lepton}, the strongest condition is given by $t=10^7{\rm [s]}$. As one can observe in the red and the green regions, the constraint at another time is all included in the yellow region. In the right panel of the figure \ref{lepton}, we show constraints at $t=10^7$[s] for two reconnection probabilities, $p^b=10^{-2}$ (the orange region) and $p^b=10^{-3}$ (the yellow region). The blue line represents the line $L_{\rm GW}=L_{\rm AB}$. Above the line, constraint is imposed by the first line of \eqref{result}. In this case, as we mentioned above, as $G\mu$ decreases the total power of radiation increases. On the other hand, below the line, the second line of \eqref{result} should be used, in which the radiation power decreases with $G\mu$. Hence, the excluded regions form a ``band-shape''. Also, in general, the constraints become strong when $k$ and $p^b$ are small.

\begin{figure}[htbp]
\begin{center}
 \includegraphics[width=.4\linewidth]{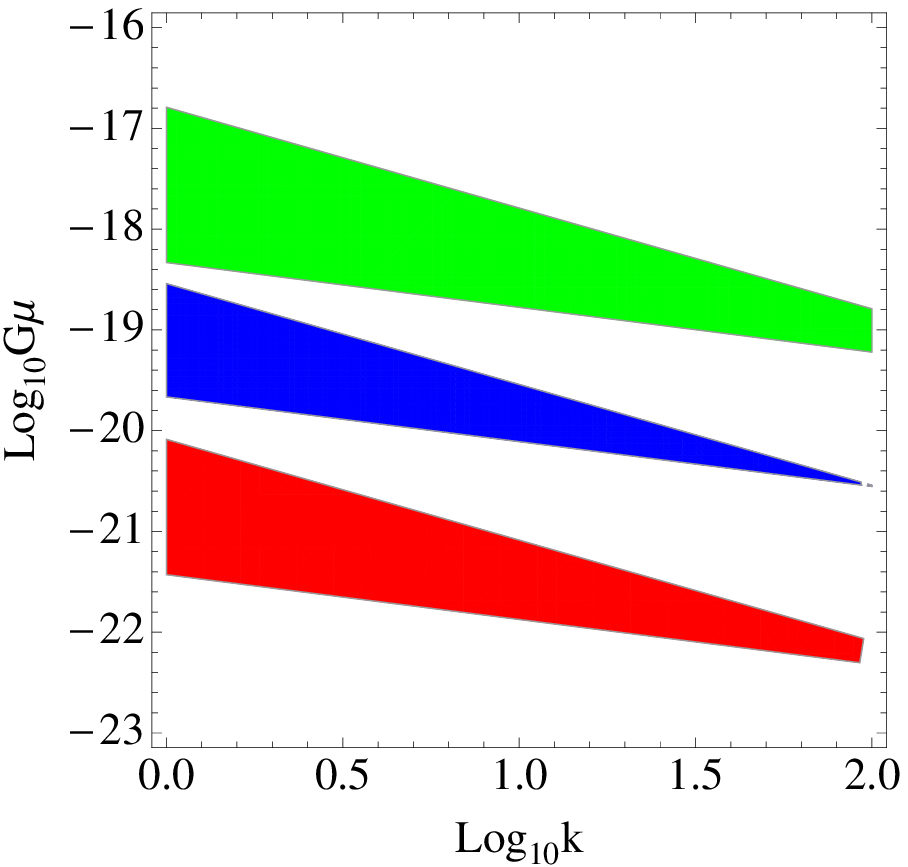} 
 \hspace{3mm}
 \includegraphics[width=.4\linewidth]{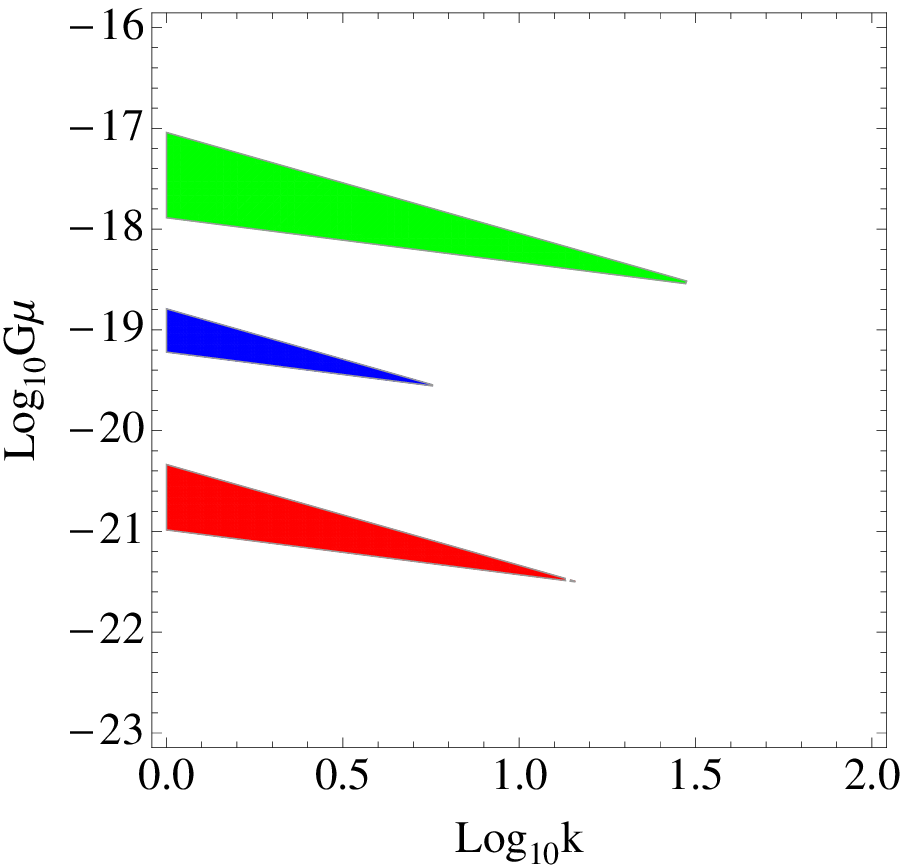} 
\vspace{-.1cm}
\caption{\sl Constraints from hadronic radiation. In the left panel, the green, blue and red regions are excluded by the conditions \eqref{hadronconstraint} for $p^b=10^{-3}$ at $t=10^{1}{\rm [s]}$, $10^4{\rm [s]}$ and $10^{8}{\rm [s]}$ respectively. The white region is allowed. We took the parameters, $\alpha=0.1$, $g_*=10.75$, $\Gamma_{\rm AB}=\Gamma_{\rm GW}=\xi=50$. In the right panel, we show the excluded regions for $p^b =10^{-2.5}$ with the same parameter choice. }
\label{hadron}
\end{center}
\end{figure}

Next, we study radiation of quarks from AB string\footnote{AB strings associated with $ \mathbb{Z}_3^{II}$ symmetry in appendix A are constrained by this condition as well as by the leptonic one. }. Right after emission, quark turns into a hadronic jet, which affects successful BBN. The constraints for the hadronic radiation can be read off from \cite{BBN} as 
\begin{align}
\label{hadronconstraint}
\frac{\rho}{s}<\left\{
\begin{array}{ll}
10^{-{17/ 2}}\left( { 1 { \rm [s]}  \over t } \right)^{5/ 2}{\rm [GeV]} & \text{for} \quad 10^{-1}\, [s]\le t < 10^0  { \rm [s]},  \\ 
10^{-{17 / 2}}{\rm \, [GeV]}  & \text{for} \quad 10^0 \,  { \rm [s]} \le t<10^2  { \rm [s]} , \\ 
10^{-6}\left({1{\rm [s]} \over t} \right)^{5/ 4}{\rm [GeV]}  & \text{for} \quad 10^2\,  { \rm [s]} \le t <10^4  { \rm [s]}, \\ 
10^{-11}{\rm [GeV]}  & \text{for} \quad 10^4\,  { \rm [s]} \le t <10^6  { \rm [s]} , \\ 
10^{-2}\left( { 1{\rm [s]}\over t} \right)^{3/ 2}{\rm [GeV]}  & \text{for} \quad 10^6\,  { \rm [s]} \le t < 10^8  { \rm [s]},  \\ 
10^{-14}{\rm [GeV]}  & \text{for} \quad 10^{8} \,  { \rm [s]}  \le t \le10^{10} { \rm [s]} . \\ 
\end{array}\right. 
\end{align}
As an illustration, we show constraints for hadronic radiation with the same parameter choice $\alpha=0.1$, $g_*=10.75$, $\Gamma_{\rm AB}=\Gamma_{\rm GW}=\xi=50$. In the left panel of the figure \ref{hadron}, the green, blue and red regions are excluded by conditions \eqref{hadronconstraint} for $p^b=10^{-3}$ at $t=10^{1}{\rm [s]}$, $10^4{\rm [s]}$ and $10^{8}{\rm [s]}$ respectively. Others are all covered by these three regions. This involved structure comes from time dependence of the line $L_{\rm GW}=L_{\rm AB}$. From the right panel, we see that the excluded regions for $p^b =10^{-2.5}$ are quite narrow. 
%%%%%%%%%%%%
%%%%%%%%%%%%

%%%%%%%%%%
\subsection{Diffuse $\gamma$-ray background }

Another stringent cosmological constraint after the recombination arises from observations of the diffuse $\gamma$-ray background. In the matter-dominated era, the number density of the loop size between $L $ and $L+d L$ is given by \cite{VS,Vach,Dufaux,dilaton}
\begin{equation}
n_L dL ={\xi \sqrt{\alpha t_{\rm eq}}   \over p^b L^{5/2} t^{2} },
\end{equation}
where $t_{\rm eq}\simeq 1.8\times 10^{12}{\rm [s]}$ is the time the matter-dominated era begins, and the factor $\sqrt{t_{\rm eq}}$ appears for continuous connection to \eqref{number density}. Hence, the total power injected from the AB string can be written as 
\begin{eqnarray}
 \dot{\rho}&=&\int^{L_{\rm max}}_{L_{\rm min}} d L\, n_L P_{\rm AB} ={2\Phi^2  \xi \sqrt{\alpha t_{\rm eq}} \sqrt{\mu} \over 5 p^b t^{2}}  \Gamma_{\rm AB} L_{\rm min}^{-5/2}+\cdots \nonumber \\
&\simeq &{2 \xi \sqrt{\alpha t_{\rm eq}} \over 5 p^b } \times  \left\{
\begin{array}{ll}
{\Phi^2 \Gamma_{\rm AB} \over  \Gamma_{\rm GW}^{5/2}} (G\mu)^{-2}M_{\rm pl}t^{-9/2} &\quad \text{for}\quad L_{\rm min}=L_{\rm GW} \\ 
 ( \Phi^{ 2}\Gamma_{AB})^{-1/ 4}(G\mu)^{9/ 8}M_{\rm pl}^{9/4} t^{-13/4} & \quad \text{for}\quad L_{\rm min}=L_{\rm AB} .
 \end{array}\right. 
\end{eqnarray}
Once electrons or quarks are produced, electromagnetic cascades are induced and the most of the energy fraction coming from the AB strings are immediately converted into a diffuse flux of $\gamma$-ray, which is constrained by the EGRET~\cite{EG} and Fermi-LAT~\cite{LAT} experiments. The constraint is given by the cascade energy density accumulated up to the present age, $t_{0}\simeq 4.3\times 10^{17}{\rm [s]}$, \cite{Vach,Dufaux,dilaton},
\begin{eqnarray}
\omega_{\rm cas} \equiv  \int^{t_0}_{t_{\rm cas}} dt \dot{\rho} {a^4(t)\over a^4(t_0)}  < 5.8\times 10^{-7} \left[ {\rm eV \over cm^3}\right], \label{gammacond}
\end{eqnarray}
%where $f_{\rm em}$ is the fraction of energy density that goes into the electromagnetic cascade and we here estimate that $f_{\rm em}\simeq 0.5$. 
where $t_{\rm cas}$ is the time electromagnetic cascades become relevant. Before this time, produced photos are absorbed by the cosmological medium. Here, we assume $t_{\rm cas}\simeq {\cal O}(10^{14\sim 15}){\rm [s]}$, following \cite{Vach,Dufaux,dilaton}.

\begin{figure}[htbp]
\begin{center}
 \includegraphics[width=.4\linewidth]{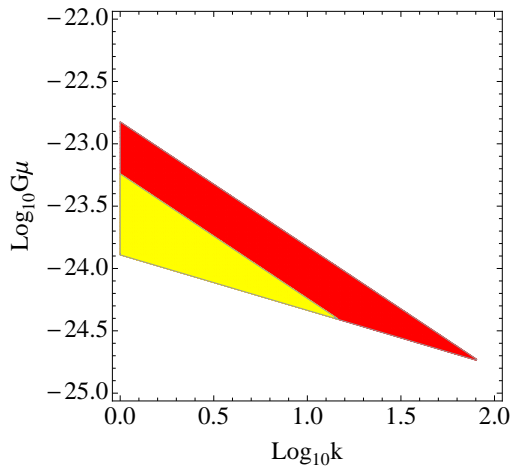}
\vspace{-.1cm}
\caption{\sl  Constraints from the diffuse $\gamma$-ray background. The red and yellow regions are excluded by the condition \eqref{gammacond} for $t_{\rm cas}=10^{14}$, $10^{15}{\rm [s]}$ with the parameters, $p^b=10^{-3}$, $\alpha=0.1$, $\Gamma_{\rm AB}=50$, $\Gamma_{\rm GW}=20$ and $\xi=50$.  }
\label{gammaray}
\end{center}
\end{figure}

Note that the minimum size of loops $L_{\rm min}$ depends on time. From the condition $L_{\rm GW}=L_{\rm AB}$, we obtain the critical time after which the gravitation wave dominates the AB radiation 
\begin{equation}
t_{\rm crit}\simeq 5.4\times 10^{-44} {\rm [s]} {\Gamma_{\rm AB}\Phi^2\over \Gamma_{\rm GW}^2}(G\mu)^{-5/2}.
\end{equation}
Hence, depending on $G\mu$ there are three cases, $t_{\rm crit}<t_{\rm cas}$, $t_{\rm cas}< t_{\rm crit}<t_0$ and $t_{\rm 0}< t_{\rm crit}$. The accumulated energy is, thus, calculated in each case separately, 
\begin{eqnarray}
\omega_{\rm cas}\simeq  K\times  \left\{
\begin{array}{lll}
 (G\mu)^{-2} \left({t_0\over t_{\rm cas}} \right)^{5/6}   &\ \   \text{for}\quad  t_{\rm crit}<t_{\rm cas} ,\\
{2 \Gamma_{\rm GW}^{5/2}\over (\Gamma_{\rm AB}\Phi^2 )^{5/4} } \left({t_{\rm crit}\over t_{0}} \right)^{5/12}(M_{\rm pl}t_0)^{5/4}  (G\mu)^{9/8}  \left[1- ({t_{\rm cas}\over t_{\rm crit}})^{5/12} \right] \\ \qquad \qquad \quad \qquad+(G\mu)^{-2} \left({t_0\over t_{\rm crit}} \right)^{5/6}  \left[ 1 -( {t_{\rm cas}\over t_{0} })^{5/6} \right]  & \ \  \text{for}\quad  t_{\rm cas}< t_{\rm crit}<t_0, \\
 {2 \Gamma_{\rm GW}^{5/2}\over (\Gamma_{\rm AB}\Phi^2 )^{5/4} } (M_{\rm pl}t_0)^{5/4}  (G\mu)^{9/8} & \ \  \text{for}\quad  t_0< t_{\rm crit},
 \end{array}\right.  \nonumber
\end{eqnarray}
where we introduced the common factor, 
\begin{equation}
K\equiv {12  \xi \sqrt{\alpha}\Gamma_{\rm AB}\Phi^2 \over 25 p^b \Gamma_{\rm GW}^{5/2}} \left({t_{\rm eq}\over t_{0}} \right)^{1/2} {M_{\rm pl}\over t_0^3 } . \nonumber
\end{equation}
Plugging this into \eqref{gammacond}, we obtain constraints for the AB radiation. In figure \ref{gammaray}, we exhibit excluded regions for $t_{\rm cas}=10^{14}, 10^{15}{\rm [s]}$ with the parameters, $p^b=10^{-3}$, $\alpha=0.1$, $\Gamma_{\rm AB}=50$, $\Gamma_{\rm GW}=20$ and $\xi=50$. Since the constraint \eqref{gammacond}  is relatively weak, we used slightly different parameter choice from the one used in section 4.1. As one can observe from the figure \ref{gammaray}, the upper boundaries of the excluded regions depend sensitively on the time $t_{\rm cas}$.  It is interesting that lower regions of $G\mu$ are excluded, which is complementary to the BBN results.

%%%%%%%%%%%
%%%%%%%%%%%

%%%%%%%%%%%%%%%
\section{Conclusions}

In this paper, we studied radiation of the standard model particles from Aharonov-Bohm strings associated with discrete gauge symmetries. In intersecting D-brane models in Type IIA string theory, Aharonov-Bohm particles correspond to the particles in the standard model, and have light masses compared to the scale of the string tension. Owing to this fact, Aharonov-Bohm radiation occurs at a drastic rate and non-trivial cosmological constraints coming from the BBN and $\gamma$-ray background were obtained. Also, small reconnection probability in string theory enhances the radiation power, so constraints shown in this paper become robust in string phenomenology. As for the case $p^b=10^{-3}$, {\it some regions} in the scale $10^{6.5} {\rm [GeV]} \lesssim \sqrt{\mu} \lesssim 10^{11} {\rm [GeV]}$, which is complementary to the LHC and future gravitational wave experiments, are excluded for small $k$. The conditions are converted into relations for the compactification scale, the string scale and the string coupling since the string tension in the D-brane model is given by $\mu = {V_{3}/ (2\pi)^4g_s l_s^5 } $. 

It would be interesting to explore further in this direction and study cosmological constraints in string theory.  Extensions to supersymmetric models\footnote{See \cite{KOO} for recent progresses on supersymmetric model buildings.} are interesting since radiation of the lightest supersymmetric particles would lead to a stringent constraint for the AB radiation. We leave it as a future work.

Finally, we would like to comment on a connection to inflations in string theories. One of the successful inflation models in string theories is KKLMMT model \cite{KKLMMT}, which  successfully stabilize moduli in the string compactification. In the particular setup, D3 and anti-D3 branes on conifold were used and at the end of the inflation, D3/anti-D3 annihilation occurs. According to \cite{KS}, after the tachyon condensation, the geometry allows to possess a discrete gauge symmetry in the IR dual field theory. The IR tip of the conifold corresponds to the gaugino condensation which breaks $U(1)_R$ symmetry in the UV theory down to $ \mathbb{Z}_2$ symmetry. A natural candidate for the Aharonov-Bohm string associated with the $ \mathbb{Z}_2$ symmetry is D1 string which can be created by the tachyon condensation of D3/anti-D3 branes. Thus, in this setup, the AB string is not diluted by the inflation but rather plays important role for probing the high energy physics. Naively we can expect that the AB string radiates the standard model particles since the annihilation of D3/anti-D3 have to reheat the standard model particle for successful inflation. So the argument shown in this paper may give a constraint for such model building. In fact, in the model KKLMMT, for successful inflation, the string scale and string coupling get a constraint which can be converted to the constraint for the tension of the D1 string, $10^{-12}\le G\mu \le 10^{-6}$ \cite{cosmicsuperstring,reviewsuperstring}. As shown in this paper, this parameter region severely constrain by the AB radiation\footnote{We would like to thank the referee for drawing our attention to this point.}. Since the detail studies of such constraints are our of  our scape, we would like to leave them as a future work.  

%%%%%%%%%%%%%%%%%%%%%
\section*{Acknowledgement}

The author is grateful to Y. Hamada, K. Kamada and T. Kobayashi for collaboration at the early stage of this project and to K. Hanaki and S. Sugimoto  for useful comments and discussions. The author would like to thank California Institute of Technology for their hospitality where the early stage of the work was done. This work is supported by Grant-in-Aid for Scientific Research from the Ministry of Education, Culture, Sports, Science and Technology, Japan (No. 25800144 and No. 25105011).

%%%%%%%%%%%%%%%%%%%%
\appendix

%%%%%%%%%%%%%%%%%%%%%%%
\setcounter{equation}{0}
\renewcommand{\theequation}{A.\arabic{equation}}
\section*{Appendix A\, Toroidal orientifold models in IIA string theory}
%%%%%%%%%

In this appendix, we discuss Type IIA orientifold models with intersecting D6-branes wrapping on three-cycles as an  illustrative example of the arguments given in the main text. Depending on the wrapping numbers of D6-branes on the three-cycles, we can engineer various discrete gauge symmetries in the four-dimensional effective theory \cite{DGS2}. Such discrete symmetries are originated from the extra $U(1)$ symmetries which are ubiquitous in realistic model buildings in string theories (see \cite{Review} for reviews). In this case, particles in the standard model naturally carry charges of the discrete symmetries. Namely, they are Aharonov-Bohm particles.

Consider four stacks of D6-branes, $N_a=3$, $N_b=N_c=N_d=1$. By exploiting the same notation as \cite{DGS2}, we refer to the four stacks as $a$, $b$, $c$, $d$.  Each has the mirror image under the orientifold projection except the stack $b$ which is on the top of the orientifold plane. Choosing the charge of the orientifold for ``$Sp$'' type, $SU(2)_L$ symmetry in the standard model is realized on $b$ as $Sp(2)=SU(2)_L$. The gauge groups in the four-dimensional effective theory are
\begin{equation}
U(3)_a\times Sp(2) \times U(1)_c\times U(1)_d.
\end{equation}
An appropriate choice of intersection numbers yields the correct numbers of the chiral fermions in the standard model. Charge assignments for each $U(1)$ symmetry is shown in the table A. One linear combination of the $U(1)$ symmetries corresponds to the hypercharge
\begin{equation}
Y\equiv {1\over 6} (Q_a-3Q_c+3Q_d), \label{Hyper}
\end{equation}
where $Q_i$'s is the generator of each $U(1)$ symmetry. 

%%%%%%%%%%%%%%%%%
\begin{table}[ht]
\begin{center}
\begin{tabular}{|c||c|c|c|c|c|c|}
\hline
 & $Q_L$ & $U_R$ & $D_R$ & $L$ & $E_R$ & $N_R$ \\
\hline
\hline
$Q_a$ & $1$ & $-1$ & $-1$ & $0$ & $0$ & $0$ \\
\hline
$Q_c$ & $0$ & $1$ & $-1$ & $0$ & $-1$ & $1$  \\
\hline
$Q_d$ & $0$ & $0$ & $0$ & $-1$ & $1$& $1$ \\
\hline
$Y$     & $1/6$ & $-2/3$ & $1/3$ & $-1/2$ & $1$ & $0$ \\
\hline
\end{tabular}
%\label{ChargeParticle}
%\caption{Table A: CAPTION}
\center{Table A: Particles in the standard model and $U(1)$ charges.}
\end{center}
\end{table}
%%%%%%%%%%%%%%%%

In the case of toroidal orientifold models, the three-cycles are constructed out of one-cycles in $ \mathbb{T}^2\times \mathbb{T}^2\times \mathbb{T}^2$. We denote the three tori as $(\mathbb{T}^2)^i$ with $i=1,2,3$, and one-cycles $a_i$, $b_j$. The intersection numbers of the one-cycles are $a_i\cdot b_j=\delta_{ij}$. Four independent pairs of three-cycles are distinguished by even or odd under the orientifold action. The cycles $\alpha_i$ (cycles $\beta_i$) are even (odd), 
\begin{eqnarray}
&&\alpha_0=a_1a_2a_3,\qquad \beta_0=b_1b_2b_3, \nonumber \\ 
&&\alpha_1=a_1b_2b_3,\qquad \beta_1=b_1a_2a_3,  \\
&&\alpha_2=b_1a_2b_3,\qquad \beta_2=a_1b_2a_3, \nonumber \\
&&\alpha_3=b_1b_2a_3,\qquad \beta_3=a_1a_2b_3. \nonumber  
\end{eqnarray}
The wrapping numbers of D6-branes on the three-cycles are given by products of the wrapping numbers of each one-cycle. Suppose that D6-branes are wrapping on three-cycles as follows,
\begin{equation}
\Gamma_A\equiv \sum_{k=0}^3 (r_A^k \alpha_k + s_A^k \beta_k),
\end{equation}
where we introduced $A=a,b,c,d$ and $r_A^k$, $s_A^k$ stand for the wrapping numbers for each three-cycle. From \eqref{Hyper}, one linear combination corresponding to the hypercharge has to be massless. To satisfy the condition, we impose
\begin{equation}
s_a^k -s_c^k +s_d^k =0,\quad {\rm for\ \ all }\ \ k.
\end{equation}
Note that as for the stack $b$ there is no $U(1)$ symmetry, $s_b^k=0$, because of $Sp(2)$ symmetry. Here we show an example having three generations of the standard model particles. For the sake of simplicity, we take the simplest case yielding non-trivial discrete gauge symmetries\footnote{We choose $n_a^{2}=0$, $n_c^{1}=n_d^{2} m_d^{3}$, $m_a^{3}=m_d^{3}=1$, $n_c^{1}=n_d^{2}=3$ and $N_g=3$ in the models studied in \cite{DGS2}.}, 
%
%%%%%%%%%%%%%%%%%%%%%%%%%%%%%%%%%%%%
\begin{table}[htb] 
\begin{center}
\begin{tabular}{|c||c|c|c|}
\hline
     &  $(n^1,m^1)$  &  $(n^2,m^2)$   & $(n^3,m^3)$ \\
\hline\hline $N_a=3$ & $(1,0)$  &  $(0, 1)$ &
 $(3,  1)$  \\
\hline $N_b=1$ &   $(0,1)$    &  $ (1,0)$  &
$(0,-1)$   \\
\hline $N_c=1$ & $(3,1)$  &
 $(1,0)$  & $(0,1)$  \\
\hline $N_d=1$ &   $(1,0)$    &  $(3,-3)$  &
$(1, 1)$   \\
\hline 
\end{tabular}
\end{center} 
\center{Table B: The wrapping numbers of D6-branes on one-cycles. }
\label{bbbbb} 
\end{table}
%%%%%%%
where $(n^{i}, m^{i})$ are the wrapping numbers on one-cycles $(a_i, b_i)$. In this case, the wrapping numbers for three-cycles are given by
\begin{equation}
s_a^2= s_c^3=-s_d^2=s_d^3=3.
\end{equation}
Others are all zero.

The Chern-Simons terms of D6-brane action contain 
\begin{equation}
S_{CS}= {1\over 2} \left(\int_{\Gamma_A } C_5\wedge F_A -\int_{\Gamma_A^*} C_5\wedge F_A \right),
\end{equation}
where $\Gamma_A^*$ stands for the mirror image of $A$. In the present example, the Chern-Simons terms lead to BF couplings,
\begin{eqnarray}
S_{CS}= 3\int_{4D}  (3F^a-F^d) \wedge  \widetilde{C}_2^2+3\int_{4D} ( F^c-F^d) \wedge \widetilde{C}_2^3, \label{BFexample}
\end{eqnarray}
where
\begin{equation}
\widetilde{C}_2^i=\int_{\beta_i}C_5.
\end{equation}
From this, we conclude that two linear combinations of $U(1)$ symmetries are massive
\begin{equation}
\tilde{Q}_I\equiv 3Q_a-Q_d,\quad \tilde{Q}_{II}\equiv Q_c-Q_d. \label{generators}
\end{equation}
The BF couplings \eqref{BFexample} indicate two discrete gauge symmetries,
\begin{equation}
\mathbb{Z}_3^I\times \mathbb{Z}_3^{II}.
\end{equation}
Aharonov-Bohm strings associated with $\mathbb{ Z}_3^I$ are string-like objects electrically coupled to $\widetilde{C}_2^2$, so in ten dimensions, it can be interpreted as a D4-brane wrapping on $\beta_2$ cycle. According to \cite{BS}, three AB strings can be annihilated with a monopole that is magnetically coupled to the gauge field, $\tilde{A}^I_{\mu}$, corresponding to the generator $\tilde{Q}_I$ in \eqref{generators}. From D6-brane point of view, the endpoint of a D4-brane can be interpreted as a magnetic monopole (for example, see \cite{GK}). In the present model, such D4-brane has a finite mass by wrapping on $\beta_2$ cycle and a one-cycle. The minimum size of one-cycle gives the lightest monopole. On the other hand, AB particles associated with $\mathbb{ Z}_3^I$ are objects electrically couple to $\tilde{A}^{I}_{\mu}$, thus they are the fundamental strings ending on $a$, $b$, $c$ branes in the context of string theory. From the table A, one can understand that $L$, $E_R$ and $N_R$ carries nontrivial charges for the $\mathbb{Z}_3^I$ symmetry\footnote{$Q_L$, $U_R$ and $D_R$ have charges of multiples of three which are trivial in the symmetry.}. Although the quarks are neutral under $ \mathbb{Z}_3^{I}$ symmetry, $U_R$ and $D_R$ have nontrivial charges for $\mathbb{Z}_3^{II}$ symmetry. 

As we explained in the main text, AB strings and AB particles have non-trivial interactions. In circling around the AB string, the AB particle picks up a non-zero holonomy. This interaction is quantum and topological, so it is relatively weak. However, direct radiation of the standard model particles with small reconnection probability gives us a cosmologically detectable size of effect.

%%%%%%%%%%%%%%%%%%%%%%%%%%%%%%%%%%%%%%%%%%%%%%
%
% Bibliography
%
%%%%%%%%%%%%%%%%%%%%%%%%%%%%%%%%%%%%%%%%%%%%%%


\begin{thebibliography}{1}


%\cite{Harigaya:2013vja}
\bibitem{Yanagida} 
  K.~Harigaya, M.~Ibe, K.~Schmitz and T.~T.~Yanagida,
  %``The Peccei-Quinn Symmetry from a Gauged Discrete R Symmetry,''
  arXiv:1308.1227 [hep-ph];  J.~L.~Evans, M.~Ibe, J.~Kehayias and T.~T.~Yanagida,
  %``Non-Anomalous Discrete R-symmetry Decrees Three Generations,''
  Phys.\ Rev.\ Lett.\  {\bf 109}, 181801 (2012)
  [arXiv:1111.2481 [hep-ph]];  M.~Asano, T.~Moroi, R.~Sato and T.~T.~Yanagida,
  %``Non-anomalous Discrete R-symmetry, Extra Matters, and Enhancement of the Lightest SUSY Higgs Mass,''
  Phys.\ Lett.\ B {\bf 705}, 337 (2011)
  [arXiv:1108.2402 [hep-ph]];  M.~Dine, F.~Takahashi and T.~T.~Yanagida,
  %``Discrete R Symmetries and Domain Walls,''
  JHEP {\bf 1007}, 003 (2010)
  [arXiv:1005.3613 [hep-th]];
  K.~Hamaguchi, Y.~Nomura and T.~Yanagida,
  %``Longlived superheavy dark matter with discrete gauge symmetries,''
  Phys.\ Rev.\ D {\bf 59}, 063507 (1999)
  [hep-ph/9809426].


  
  
 \bibitem{examples}
  H.~K.~Dreiner, C.~Luhn, M.~Thormeier,
%  ``What is the discrete gauge symmetry of the MSSM?,''
  Phys.\ Rev.\  {\bf D73 } (2006)  075007.
  [hep-ph/0512163];
  R.~N.~Mohapatra, M.~Ratz,
 % ``Gauged Discrete Symmetries and Proton Stability,''
  Phys.\ Rev.\  {\bf D76 } (2007)  095003.
  [arXiv:0707.4070 [hep-ph]];
  T.~Araki, T.~Kobayashi, J.~Kubo, S.~Ramos-Sanchez, M.~Ratz and P.~K.~S.~Vaudrevange,
%  ``(Non-)Abelian discrete anomalies,''
  Nucl.\ Phys.\  B {\bf 805} (2008) 124
  [arXiv:0805.0207 [hep-th]];
  H.~M.~Lee, S.~Raby, M.~Ratz, G.~G.~Ross, R.~Schieren, K.~Schmidt-Hoberg, P.~K.~S.~Vaudrevange,
%  ``A unique $Z_4^R$ symmetry for the MSSM,''
  Phys.\ Lett.\  {\bf B694}, 491-495 (2011).
  [arXiv:1009.0905 [hep-ph]];
  R.~Kappl, B.~Petersen, S.~Raby, M.~Ratz, R.~Schieren, P.~K.~S.~Vaudrevange,
 % ``String-derived MSSM vacua with residual R symmetries,''
  Nucl.\ Phys.\  {\bf B847 } (2011)  325-349.
  [arXiv:1012.4574 [hep-th]];
  A.~Font, L.~E.~Ib\'a\~nez, F.~Quevedo,
%  ``Does Proton Stability Imply the Existence of an Extra Z0?,''
  Phys.\ Lett.\  {\bf B228 } (1989)  79.

%\cite{Ishimori:2010au}
\bibitem{Ishimori} 
  H.~Ishimori, T.~Kobayashi, H.~Ohki, Y.~Shimizu, H.~Okada and M.~Tanimoto,
  %``Non-Abelian Discrete Symmetries in Particle Physics,''
  Prog.\ Theor.\ Phys.\ Suppl.\  {\bf 183}, 1 (2010)
  [arXiv:1003.3552 [hep-th]].
  %%CITATION = ARXIV:1003.3552;%%
  %230 citations counted in INSPIRE as of 07 Oct 2013


%\cite{Banks:2010zn}
\bibitem{BS} 
  T.~Banks and N.~Seiberg,
%  ``Symmetries and Strings in Field Theory and Gravity,''
  Phys.\ Rev.\ D {\bf 83}, 084019 (2011)
  [arXiv:1011.5120 [hep-th]].
  %%CITATION = ARXIV:1011.5120;%%
  %58 citations counted in INSPIRE as of 12 Jul 2013
%
%%\cite{Hellerman:2010fv}
%\bibitem{HS} 
%  S.~Hellerman and E.~Sharpe,
%  %``Sums over topological sectors and quantization of Fayet-Iliopoulos parameters,''
%  Adv.\ Theor.\ Math.\ Phys.\  {\bf 15}, 1141 (2011)
%  [arXiv:1012.5999 [hep-th]].
%  %%CITATION = ARXIV:1012.5999;%%
%  %23 citations counted in INSPIRE as of 07 Oct 2013



%\cite{Banks:1988yz}
\bibitem{OldNoGlobal}
  T.~Banks, L.~J.~Dixon,
%  ``Constraints on String Vacua with Space-Time Supersymmetry,''
  Nucl.\ Phys.\  {\bf B307 } (1988)  93-108;   L.~F.~Abbott, M.~B.~Wise,
 % ``Wormholes And Global Symmetries,''
  Nucl.\ Phys.\  {\bf B325 } (1989)  687;  S.~R.~Coleman, K.~-M.~Lee,
  ``Wormholes Made Without Massless Matter Fields,''
  Nucl.\ Phys.\  {\bf B329 } (1990)  387.
  
  



%\cite{Kallosh:1995hi}
\bibitem{NewNoGlobal}
  R.~Kallosh, A.~D.~Linde, D.~A.~Linde, L.~Susskind,
%  ``Gravity and global symmetries,''
  Phys.\ Rev.\  {\bf D52 } (1995)  912-935.
  [hep-th/9502069].




\bibitem{OldAB}
   L.~M.~Krauss, F.~Wilczek,
%  ``Discrete Gauge Symmetry in Continuum Theories,''
  Phys.\ Rev.\ Lett.\  {\bf 62 } (1989)  1221;  M.~G.~Alford, J.~March-Russell and F.~Wilczek,
%  ``Discrete Quantum Hair On Black Holes And The Nonabelian Aharonov-bohm Effect,''
  Nucl.\ Phys.\ B {\bf 337} (1990) 695; J.~Preskill, L.~M.~Krauss,
%  ``Local Discrete Symmetry And Quantum Mechanical Hair,''
  Nucl.\ Phys.\  {\bf B341 } (1990)  50-100.


%\cite{Aharonov:1959fk}
\bibitem{Original}
  Y.~Aharonov and D.~Bohm,
%  ``Significance of electromagnetic potentials in the quantum theory,''
  Phys.\ Rev.\  {\bf 115}, 485 (1959).
  %%CITATION = PHRVA,115,485;%%



%\cite{Bizet:2013wha}
\bibitem{DSmany} 
  N.~G.~C.~Bizet, T.~Kobayashi, D.~K.~M.~Pena, S.~L.~Parameswaran, M.~Schmitz and I.~Zavala,
%  ``Discrete R-symmetries and Anomaly Universality in Heterotic Orbifolds,''
  arXiv:1308.5669 [hep-th],  JHEP {\bf 1305}, 076 (2013)
 [arXiv:1301.2322 [hep-th]];
 %%CITATION = ARXIV:1301.2322;%%
    H.~P.~Nilles, S.~Ramos-Sanchez, M.~Ratz and P.~K.~S.~Vaudrevange,
  %``A note on discrete R symmetries in Z6-II orbifolds with Wilson lines,''
  arXiv:1308.3435 [hep-th];
  %%CITATION = ARXIV:1308.3435;%%
  %1 citations counted in INSPIRE as of 07 Oct 2013
  J.~E.~Kim,
  %``Abelian discrete symmetries Z_{N} and Z_{nR} from string orbifolds,''
  arXiv:1308.0344 [hep-th];
  %%CITATION = ARXIV:1308.0344;%%
  %%CITATION = ARXIV:1306.1284;%%
  %4 citations counted in INSPIRE as of 07 Oct 2013 
  P.~Anastasopoulos, M.~Cvetic, R.~Richter and P.~K.~S.~Vaudrevange,
  %``String Constraints on Discrete Symmetries in MSSM Type II Quivers,''
  JHEP {\bf 1303}, 011 (2013)
  [arXiv:1211.1017 [hep-th]].
  %%CITATION = ARXIV:1211.1017;%%
  %8 citations counted in INSPIRE as of 07 Oct 2013


%%%%%%Recent Study Discerte Gauge symmetry in String %%%%%%

%\cite{Camara:2011jg}
\bibitem{DGS1}
  P.~G.~Camara, L.~E.~Ibanez and F.~Marchesano,
%  ``RR photons,''
  JHEP {\bf 1109} (2011) 110
  [arXiv:1106.0060 [hep-th]].
  %%CITATION = ARXIV:1106.0060;%%

%\cite{BerasaluceGonzalez:2011wy}
\bibitem{DGS2}
  M.~Berasaluce-Gonzalez, L.~E.~Ibanez, P.~Soler and A.~M.~Uranga,
 % ``Discrete gauge symmetries in D-brane models,''
  JHEP {\bf 1112} (2011) 113
  [arXiv:1106.4169 [hep-th]].
  %%CITATION = ARXIV:1106.4169;%%

%\cite{Ibanez:2012wg}
\bibitem{DGS3}
  L.~E.~Ibanez, A.~N.~Schellekens and A.~M.~Uranga,
 % ``Discrete Gauge Symmetries in Discrete MSSM-like Orientifolds,''
arXiv:1205.5364 [hep-th].
%%CITATION = ARXIV:1205.5364;%


%\cite{BerasaluceGonzalez:2012zn}
\bibitem{DGS4} 
  M.~Berasaluce-Gonzalez, P.~G.~Camara, F.~Marchesano and A.~M.~Uranga,
%  ``${\bf Z}_p$ charged branes in flux compactifications,''
  JHEP {\bf 1304}, 138 (2013)
  [arXiv:1211.5317 [hep-th]].
  %%CITATION = ARXIV:1211.5317;%%
  %4 citations counted in INSPIRE as of 03 Oct 2013

%\cite{BerasaluceGonzalez:2012vb}
\bibitem{DGS5} 
  M.~Berasaluce-Gonzalez, P.~G.~Camara, F.~Marchesano, D.~Regalado and A.~M.~Uranga,
  %``Non-Abelian discrete gauge symmetries in 4d string models,''
  JHEP {\bf 1209}, 059 (2012)
  [arXiv:1206.2383 [hep-th]].
  %%CITATION = ARXIV:1206.2383;%%
  %19 citations counted in INSPIRE as of 07 Oct 2013
    F.~Marchesano, D.~Regalado and L.~Vazquez-Mercado,
  %``Discrete flavor symmetries in D-brane models,''
  JHEP {\bf 1309}, 028 (2013)
  [arXiv:1306.1284 [hep-th]];
 H.~Abe, K.~-S.~Choi, T.~Kobayashi and H.~Ohki,
 %``Non-Abelian Discrete Flavor Symmetries from Magnetized/Intersecting Brane Models,''
 Nucl.\ Phys.\ B {\bf 820}, 317 (2009)
 [arXiv:0904.2631 [hep-ph]];
 %%CITATION = ARXIV:0904.2631;%%
 T.~Kobayashi, S.~Raby and R.~-J.~Zhang,
 %``Searching for realistic 4d string models with a Pati-Salam
 %symmetry: Orbifold grand unified theories from heterotic string
 %compactification 
 %on a Z(6) orbifold,''
 Nucl.\ Phys.\ B {\bf 704}, 3 (2005)
 [hep-ph/0409098];
 %%CITATION = HEP-PH/0409098;%%
 T.~Kobayashi, H.~P.~Nilles, F.~Ploger, S.~Raby and M.~Ratz,
 %``Stringy origin of non-Abelian discrete flavor symmetries,''
 Nucl.\ Phys.\ B {\bf 768}, 135 (2007)
 [hep-ph/0611020].
 %%CITATION = HEP-PH/0611020;%%



%\cite{Alford:1988sj}
\bibitem{AW}
  M.~G.~Alford and F.~Wilczek,
 % ``Aharonov-Bohm Interaction of Cosmic Strings with Matter,''
  Phys.\ Rev.\ Lett.\  {\bf 62}, 1071 (1989).
  %%CITATION = PRLTA,62,1071;%%





%\cite{JonesSmith:2009ti}
\bibitem{AB1} 
  K.~Jones-Smith, H.~Mathur and T.~Vachaspati,
%  ``Aharonov-Bohm Radiation,''
  Phys.\ Rev.\ D {\bf 81}, 043503 (2010)
  [arXiv:0911.0682 [hep-th]].
  %%CITATION = ARXIV:0911.0682;%%


%\cite{Chu:2010zzb}
\bibitem{AB2} 
  Y.~-Z.~Chu, H.~Mathur and T.~Vachaspati,
%  ``Aharonov-Bohm Radiation of Fermions,''
  Phys.\ Rev.\ D {\bf 82}, 063515 (2010)
  [arXiv:1003.0674 [hep-th]].
  %%CITATION = ARXIV:1003.0674;%%


%\cite{Copeland:2003bj}
\bibitem{cosmicsuperstring} 
  E.~J.~Copeland, R.~C.~Myers and J.~Polchinski,
  %``Cosmic F and D strings,''
  JHEP {\bf 0406}, 013 (2004)
  [hep-th/0312067];
  %%CITATION = HEP-TH/0312067;%%
  %353 citations counted in INSPIRE as of 08 Oct 2013
  J.~Polchinski,
  %``Open heterotic strings,''
  JHEP {\bf 0609}, 082 (2006)
  [hep-th/0510033];
  %%CITATION = HEP-TH/0510033;%%
  %32 citations counted in INSPIRE as of 08 Oct 2013
  J.~Polchinski and J.~V.~Rocha,
  %``Cosmic string structure at the gravitational radiation scale,''
  Phys.\ Rev.\ D {\bf 75}, 123503 (2007)
  [gr-qc/0702055 [GR-QC]].
  %%CITATION = GR-QC/0702055;%%
  %45 citations counted in INSPIRE as of 08 Oct 2013


  %\cite{Polchinski:2004ia}
\bibitem{reviewsuperstring} 
  J.~Polchinski,
  %``Introduction to cosmic F- and D-strings,''
  hep-th/0412244;
  %%CITATION = HEP-TH/0412244;%%
  %172 citations counted in INSPIRE as of 08 Oct 2013} 
  E.~J.~Copeland and T.~W.~B.~Kibble,
  %``Cosmic Strings and Superstrings,''
  Proc.\ Roy.\ Soc.\ Lond.\ A {\bf 466}, 623 (2010)
  [arXiv:0911.1345 [hep-th]].
  %%CITATION = ARXIV:0911.1345;%%
  %65 citations counted in INSPIRE as of 08 Oct 2013

\bibitem{VS}
 A. Vilenkin and E.P.S. Shellard,
% ``Cosmic Strings and Other Topological Defects'',
 Cambridge University Press (1994);
  M.~B.~Hindmarsh and T.~W.~B.~Kibble,
  %``Cosmic strings,''
  Rept.\ Prog.\ Phys.\  {\bf 58}, 477 (1995)
  [hep-ph/9411342].
  %%CITATION = HEP-PH/9411342;%%
  %598 citations counted in INSPIRE as of 08 Oct 2013



%\cite{Vachaspati:2009kq}
\bibitem{Vach} 
  T.~Vachaspati,
  %``Cosmic Rays from Cosmic Strings with Condensates,''
  Phys.\ Rev.\ D {\bf 81}, 043531 (2010)
  [arXiv:0911.2655 [astro-ph.CO]].
  %%CITATION = ARXIV:0911.2655;%%
  %15 citations counted in INSPIRE as of 14 Oct 2013


 %\cite{Dufaux:2011da}
\bibitem{Dufaux} 
  J.~-F.~Dufaux,
%  ``Constraints on Cosmic Super-Strings from Kaluza-Klein Emission,''
  Phys.\ Rev.\ Lett.\  {\bf 109}, 011601 (2012)
  [arXiv:1109.5121 [hep-th]]; 
  %%CITATION = ARXIV:1109.5121;%%
  %6 citations counted in INSPIRE as of 16 Jul 2013
 %\cite{Dufaux:2012np}
  J.~-F.~Dufaux,
 % ``Cosmic Super-Strings and Kaluza-Klein Modes,''
  JCAP {\bf 1209}, 022 (2012)
  [arXiv:1201.4850 [hep-th]].
  %%CITATION = ARXIV:1201.4850;%%
  %7 citations counted in INSPIRE as of 16 Jul 2013
 
 %\cite{Berezinsky:2011cp}
\bibitem{dilaton} 
  V.~Berezinsky, E.~Sabancilar and A.~Vilenkin,
 % ``Extremely High Energy Neutrinos from Cosmic Strings,''
  Phys.\ Rev.\ D {\bf 84}, 085006 (2011)
  [arXiv:1108.2509 [astro-ph.CO]].
  %%CITATION = ARXIV:1108.2509;%%
  %12 citations counted in INSPIRE as of 03 Oct 2013
 
 
 
 
%\cite{Jackson:2004zg}
\bibitem{JonsPol} 
  M.~G.~Jackson, N.~T.~Jones and J.~Polchinski,
  %``Collisions of cosmic F and D-strings,''
  JHEP {\bf 0510}, 013 (2005)
  [hep-th/0405229].
  %%CITATION = HEP-TH/0405229;%%
  %184 citations counted in INSPIRE as of 03 Oct 2013 
 
 %\cite{Damour:2004kw}
\bibitem{DV1} 
  T.~Damour and A.~Vilenkin,
%  ``Gravitational radiation from cosmic (super)strings: Bursts, stochastic background, and observational windows,''
  Phys.\ Rev.\ D {\bf 71}, 063510 (2005)
  [hep-th/0410222].
  %%CITATION = HEP-TH/0410222;%%
  %211 citations counted in INSPIRE as of 03 Oct 2013
 
 %\cite{Jones:2003da}
\bibitem{Tye1} 
  N.~T.~Jones, H.~Stoica and S.~H.~H.~Tye,
 % ``The Production, spectrum and evolution of cosmic strings in brane inflation,''
  Phys.\ Lett.\ B {\bf 563}, 6 (2003)
  [hep-th/0303269].
  %%CITATION = HEP-TH/0303269;%%
  %173 citations counted in INSPIRE as of 03 Oct 2013
 
 %\cite{Avgoustidis:2005nv}
\bibitem{AvS} 
  A.~Avgoustidis and E.~P.~S.~Shellard,
 % ``Effect of reconnection probability on cosmic (super)string network density,''
  Phys.\ Rev.\ D {\bf 73}, 041301 (2006)
  [astro-ph/0512582].
  %%CITATION = ASTRO-PH/0512582;%%
  %56 citations counted in INSPIRE as of 03 Oct 2013
 
 
 

 
%%%%%%%%%%%
\bibitem{sim1}
  V.~Vanchurin, K.~D.~Olum and A.~Vilenkin,
 % ``Scaling of cosmic string loops,''
  Phys.\ Rev.\  D {\bf 74}, 063527 (2006)
  [arXiv:gr-qc/0511159].
  K.~D.~Olum and V.~Vanchurin,
%  ``Cosmic string loops in the expanding Universe,''
  Phys.\ Rev.\  D {\bf 75}, 063521 (2007)
  [arXiv:astro-ph/0610419]. 

%%%%%%%%%%%
\bibitem{sim2}
  J.~J.~Blanco-Pillado, K.~D.~Olum and B.~Shlaer,
%  ``Large parallel cosmic string simulations: New results on loop production,''
  arXiv:1101.5173 [astro-ph.CO].

 



%\cite{Kibble:1980mv}
\bibitem{Kibble} 
 T.~W.~B.~Kibble,
% ``Some Implications of a Cosmological Phase Transition,''  
Phys.\ Rept.\  {\bf 67}, 183 (1980);
%%CITATION = PRPLC,67,183;%%
W. H. Zurek,
%``Cosmological experiments in superfluid helium?,"
Nature {\bf 317}, 505-508 (1985);
%``Cosmological experiments in condensed matter systems,"
Phys. Rep. {\bf 276}, 177-221 (1996).

 

 %\cite{Kawasaki:2004qu}
\bibitem{BBN}
  M.~Kawasaki, K.~Kohri and T.~Moroi,
%  ``Big-Bang nucleosynthesis and hadronic decay of long-lived massive particles,''
  Phys.\ Rev.\ D {\bf 71} (2005) 083502
  [astro-ph/0408426].
  %%CITATION = ASTRO-PH/0408426;%%
  %474 citations counted in INSPIRE as of 07 Jun 2013 


   %\cite{Kawasaki:2007mk}
\bibitem{Nakayama} 
  M.~Kawasaki, K.~Nakayama and M.~Senami,
  %``Cosmological implications of supersymmetric axion models,''
  JCAP {\bf 0803}, 009 (2008)
  [arXiv:0711.3083 [hep-ph]].
  %%CITATION = ARXIV:0711.3083;%%
  %57 citations counted in INSPIRE as of 09 Oct 2013
 
%\cite{Hamada:2012fr}
\bibitem{Raxion1} 
  Y.~Hamada, K.~Kamada, T.~Kobayashi and Y.~Ookouchi,
%  ``Cosmological constraints on spontaneous R-symmetry breaking models,''
  JCAP {\bf 1304}, 043 (2013)
  [arXiv:1211.5662 [hep-ph]]; %  ``More on cosmological constraints on spontaneous R-symmetry breaking models,''
  arXiv:1310.0118 [hep-ph].
  %%CITATION = ARXIV:1310.0118;%%
  %%CITATION = ARXIV:1211.5662;%%
  %2 citations counted in INSPIRE as of 04 Oct 2013 
 
%\cite{Kawasaki:2013ae}
\bibitem{Kawasaki} 
  M.~Kawasaki and K.~Nakayama,
  %``Axions : Theory and Cosmological Role,''
  arXiv:1301.1123 [hep-ph].
  %%CITATION = ARXIV:1301.1123;%%
  %13 citations counted in INSPIRE as of 09 Oct 2013 
 
 
 
 \bibitem{Gammaray}
  J.~R.~Ellis, G.~B.~Gelmini, J.~L.~Lopez, D.~V.~Nanopoulos and S.~Sarkar,
  ``Astrophysical Constraints On Massive Unstable Neutral Relic Particles,''
  Nucl.\ Phys.\  B {\bf 373}, 399 (1992);  V.~S.~Berezinsky,
  ``Neutrino astronomy and massive longlived particles from big bang,''
  Nucl.\ Phys.\  B {\bf 380}, 478 (1992).

 
\bibitem{EG}
  P.~Sreekumar {\it et al.}  [EGRET Collaboration],
  ``EGRET observations of the extragalactic gamma ray emission,''
  Astrophys.\ J.\  {\bf 494}, 523 (1998)
  [arXiv:astro-ph/9709257].

\bibitem{LAT}
  A.~A.~Abdo {\it et al.}  [The Fermi-LAT collaboration],
  ``The Spectrum of the Isotropic Diffuse Gamma-Ray Emission Derived From
  First-Year Fermi Large Area Telescope Data,''
  Phys.\ Rev.\ Lett.\  {\bf 104}, 101101 (2010)
  [arXiv:1002.3603 [astro-ph.HE]].



 %\cite{Kitano:2010fa}
\bibitem{KOO} 
  R.~Kitano, H.~Ooguri and Y.~Ookouchi,
  %``Supersymmetry Breaking and Gauge Mediation,''
  Ann.\ Rev.\ Nucl.\ Part.\ Sci.\  {\bf 60}, 491 (2010)
  [arXiv:1001.4535 [hep-th]].
  %%CITATION = ARXIV:1001.4535;%%
  %40 citations counted in INSPIRE as of 08 Oct 2013



\bibitem{Review}
  R.~Blumenhagen, M.~Cvetic, P.~Langacker, G.~Shiu,
%  ``Toward realistic intersecting D-brane models,''
  Ann.\ Rev.\ Nucl.\ Part.\ Sci.\  {\bf 55 } (2005)  71-139.
  [hep-th/0502005];
  R.~Blumenhagen, B.~Kors, D.~Lust, S.~Stieberger,
 % ``Four-dimensional String Compactifications with D-Branes, Orientifolds and Fluxes,''
  Phys.\ Rept.\  {\bf 445 } (2007)  1-193.
  [hep-th/0610327];
  L.~E.~Ib\'a\~nez, A.~M.~Uranga,
 % ``String Theory and Particle Physics: An Introduction to String Phenomenology,''
  Cambridge University Press (2012).
   
 
%\cite{Giveon:1998sr}
\bibitem{GK} 
  A.~Giveon and D.~Kutasov,
  %``Brane dynamics and gauge theory,''
  Rev.\ Mod.\ Phys.\  {\bf 71}, 983 (1999)
  [hep-th/9802067];
  %%CITATION = HEP-TH/9802067;%%
  %339 citations counted in INSPIRE as of 06 Oct 2013
  J.~Polchinski,
  %``String theory. Vol. 2: Superstring theory and beyond,''
  Cambridge University Press. (1998) 
  
  
  %\cite{Kachru:2003sx}
\bibitem{KKLMMT} 
  S.~Kachru, R.~Kallosh, A.~D.~Linde, J.~M.~Maldacena, L.~P.~McAllister and S.~P.~Trivedi,
  %``Towards inflation in string theory,''
  JCAP {\bf 0310}, 013 (2003)
  [hep-th/0308055].
  %%CITATION = HEP-TH/0308055;%%
  %820 citations counted in INSPIRE as of 09 Dec 2013
  
  %\cite{Klebanov:2000hb}
\bibitem{KS} 
  I.~R.~Klebanov and M.~J.~Strassler,
  %``Supergravity and a confining gauge theory: Duality cascades and chi SB resolution of naked singularities,''
  JHEP {\bf 0008}, 052 (2000)
  [hep-th/0007191].
  %%CITATION = HEP-TH/0007191;%%
  %1237 citations counted in INSPIRE as of 09 Dec 2013
  
\end{thebibliography}
\end{document}